\documentclass[draft]{agujournal2019}
\usepackage{url} %this package should fix any errors with URLs in refs.
\usepackage{lineno}
\usepackage[inline]{trackchanges} %for better track changes. finalnew option will compile document with changes incorporated.
\usepackage{soul}
\draftfalse
\journalname{Geophysical Research Letters}

\begin{document}

%% ------------------------------------------------------------------------ %%
%  Title
%
%% ------------------------------------------------------------------------ %%

\title{The Effect of Ocean Salinity on Climate and its Implications for Earth’s Habitability}

%% ------------------------------------------------------------------------ %%
%
%  AUTHORS AND AFFILIATIONS
%
%% ------------------------------------------------------------------------ %%

\authors{Stephanie L. Olson\affil{1}, Malte F. Jansen\affil{2}, Dorian S. Abbot\affil{2}, Itay Halevy\affil{3}, Colin Goldblatt\affil{4}}

\affiliation{1}{Department of Earth, Atmospheric, and Planetary Science, Purdue University}
\affiliation{2}{Department of the Geophysical Sciences, University of Chicago}
\affiliation{3}{Department of Earth and Planetary Sciences, Weizmann Institute of Science}
\affiliation{2}{School of Earth and Ocean Sciences, University of Victoria}

%% Corresponding Author:
% Corresponding author mailing address and e-mail address:
\correspondingauthor{Stephanie Olson}{stephanieolson@purdue.edu}

%% Keypoints
\begin{keypoints}
\item Saltier oceans result in warmer climates with less sea ice
\item Warming with increasing salinity is strongly affected by ocean dynamics
\item A saltier ocean may have helped keep early Earth warm when the Sun was less luminous
\end{keypoints}

%% ------------------------------------------------------------------------ %%
%
%  ABSTRACT and PLAIN LANGUAGE SUMMARY
%
%% ------------------------------------------------------------------------ %%

\begin{abstract}
The influence of atmospheric composition on the climates of present-day and early Earth has been studied extensively, but the role of ocean composition has received less attention. We use the ROCKE-3D ocean-atmosphere general circulation model to investigate the response of Earth’s present-day and Archean climate system to low vs. high ocean salinity. We find that saltier oceans yield warmer climates in large part due to changes in ocean dynamics. Increasing ocean salinity from 20 g/kg to 50 g/kg results in a 71\% reduction in sea ice cover in our present-day Earth scenario. This same salinity change also halves the pCO$_2$ threshold at which Snowball glaciation occurs in our Archean scenarios. In combination with higher levels of greenhouse gases such as CO$_2$ and CH$_4$, a saltier ocean may allow for a warm Archean Earth with only seasonal ice at the poles despite receiving ~20\% less energy from the Sun. 
\end{abstract}

\section*{Plain Language Summary}
The composition of the atmosphere, especially the abundance of greenhouse gases, famously influences Earth's climate system. We use a climate model to show that the composition of the ocean can also have a major impact on surface temperature and ice cover. We focus specifically on the amount of salt dissolved in seawater, and we find that saltier oceans tend to result in warmer climates. These effects are modest today, but salt may be a key ingredient for early Earth habitability in the distant past when the Sun was less bright.

%% ------------------------------------------------------------------------ %%
%
%  TEXT
%
%% ------------------------------------------------------------------------ %%

\section{Introduction}
The evolution of Earth’s climate system is intimately linked to the chemical evolution of Earth’s ocean. For example, the scarcity of dissolved electron acceptors such as oxygen and sulfate prior to Earth’s Great Oxidation Event may have allowed large fluxes of biogenic methane from the ocean to the atmosphere \cite{catling_biogenic_2001, pavlov_methane-rich_2003}. This methane may have been important for climate stability in light of the Faint Young Sun \cite{feulner_faint_2012}, and the subsequent loss of this methane flux with the progressive oxidation of Earth’s surface environments may have triggered snowball glaciation in the Paleo- and Neoproterozoic \cite{kasting_methane_2005}. Higher silica concentrations prior to the evolution of diatoms may also affect climate by promoting clay formation via reverse weathering, a process that is a net source of CO$_2$ \cite{isson_reverse_2018}. Moreover, gas solubility decreases as the concentration of dissolved ions increases, with implications for the partitioning of greenhouse gases between the ocean and the atmosphere if the salinity of the ocean has changed through time. 

The climate impacts of ocean composition are not simply limited to its interactions with greenhouse gases. The density of seawater increases with increasing salinity, and the salt content of seawater simultaneously alters the relationship between temperature and density. Salt also depresses the freezing point of seawater and may inhibit sea ice formation in salty oceans \cite{fofonoff_algorithms_1983}. Freshwater at the surface is most dense at 4 $^{\circ}$C, well above its freezing point of 0 $^{\circ}$C. The temperature of maximum density decreases with increasing salinity such that water with salinity exceeding 24 g/kg monotonically increases in density as it approaches its freezing point. Present-day seawater with a salinity of 35 g/kg freezes (and is most dense) at -1.9 $^{\circ}$C, and saltier oceans freeze at progressively lower temperatures. In combination, these three density effects may profoundly affect the density structure of the ocean, its circulation, and ocean heat transport to high latitudes with consequences for sea ice formation \cite{cael_oceans_2017, cullum_importance_2016}. Even small differences in sea ice formation may yield significant climate differences through interaction with the positive ice-albedo feedback. 

Sodium (Na$^+$) and chlorine (Cl$^-$) are the primary ions contributing to ocean salinity today. The residence times of Na$^+$ and Cl$^-$ ions in the ocean are 80 Myr and 98 Myr, respectively \cite{emerson_chemical_2008}, much shorter than the age of the Earth. There is thus every reason to expect that salinity has changed over Earth history. Such changes are possible in response to any phenomenon that changes either the amount of water in the ocean or ion fluxes into and out of the ocean. For example, water loss to space or subduction of hydrated minerals may change the amount of water in the ocean \cite{korenaga_global_2017, pope_isotope_2012}. Meanwhile, the amount of salt may change due to changes to hydrothermal or weathering inputs of ions to the ocean as Earth cooled and the continents grew. The amount of salt in the ocean may also change due to variable removal of salt from the ocean through evaporite precipitation \cite{hay_evaporites_2006}, perhaps due to dramatically differing continental configuration or climate conditions. Such changes may affect not just the total amount of dissolved ions in seawater, but also the relative abundances of specific ions. For example, Fe$^{2}$ emanating from hydrothermal systems would have been relatively abundant in an anoxic Archean ocean whereas SO$_{4}^{2-}$ would have been relatively scarce as the consequence of limited continental exposure and oxidative weathering \cite{albarede_chemical_2020}. The salinity evolution of Earth’s ocean is not yet well constrained, but constant salinity through time would be a notable coincidence or imply some currently unknown feedback. Climate models that implicitly assume present-day salinity may thus yield misleading views of Earth’s climate history. We address this gap here. 

We begin by exploring the response of present-day Earth climate to low vs. high ocean salinity. We then probe the relative contribution of ocean dynamics and freezing point depression to warming by conducting sensitivity experiments in which each effect has been disabled. We conclude by investigating the potential importance of ocean salinity on early Earth. We focus in particular on the Archean eon when Earth received ~20\% less energy from the Sun (Gough, 1981), but there is nonetheless clear evidence for liquid water, a productive marine biosphere, and biological impacts on oceanic and atmospheric chemistry \cite{lyons_rise_2014}.

\section{Model Description}
We use ROCKE-3D \cite{way_resolving_2017}, a fully coupled ocean-atmosphere GCM that includes a thermodynamic-dynamic sea ice model, to simulate the climates of Earth-like planets with oceans of differing salinity. Our simulations have a latitude-longitude resolution of 4x5 degrees, and the atmosphere includes 40 vertical layers that extend up to 0.1 mbar while the ocean includes 10 depth layers. Our present-day simulations use Earth’s current continental configuration, but we simulate the Archean Earth as an aquaplanet lacking exposed land given the prevailing view that there was limited land area and considerable uncertainty regarding its distribution. By some estimates, exposed land comprised only 7\% of Earth’s surface in the Archean, for which an aquaplanet is a reasonable approximation \cite{albarede_chemical_2020}. Both continent scenarios assume a relatively shallow, flat-bottomed ocean that is uniformly 1360 m deep, which limits computational expense and dramatically improves runtimes. This simplification nonetheless reproduces key aspects of present-day Earth’s ocean circulation, including the northward flow of warm surface waters, sinking of dense waters at high latitudes, and southward flow of cold water at depth in the Atlantic (Figure A1).   

 \begin{table}
 \caption{Summary of our Archean model configuration. Planetary parameters not specified here are fixed at their present-day values}
 \centering
 \begin{tabular}{l l l}
 \hline
  Parameter  & Value & Reference  \\
 \hline
   Insolation  & 1106.81 Wm$^{-2}$  & Gough, 1981  \\
   Day length  & 17.25 hrs & Williams, 2000 \\
   Total Atmospheric pressure & 0.5 bar & Som et al., 2016 \\
  pCH$_4$  & 500 $\mu$bar & Izon et al., 2017, Reinhard et. al., 2020 \\
  Continental distribution  & aquaplanet & Albarede et al., 2020 \\
  Ocean depth & 1360 m &  \\
  pCO$_2$  & 10-100x PIL & Hessler et al., 2004; Dreise et al., 2011 \\
 Salinity & 20-50 g/kg & Albarede et al., 2020 \\
 \hline
 \end{tabular}
 \end{table}

For each continent configuration, we simulate the steady-state climate for three global-mean ocean salinities: 20 g/kg, 35 g/kg (modern), and 50 g/kg. ROCKE-3D calculates the effects of salinity on (1) seawater density, ocean circulation patterns, and associated heat transport; and (2) freezing point depression \cite{fofonoff_algorithms_1983}. ROCKE-3D is initialized with globally homogeneous salt distributions, but ROCKE-3D conserves mass and concentrates salt in response to sea ice formation as well as evaporation and precipitation, resulting in spatial and temporal variations in salinity in our simulations that include a dynamic ocean. 

We use three complementary model configurations to probe the relative significance of dynamic effects vs. freezing point depression for climate:
 
\begin{enumerate}
\item a fully coupled ocean-atmosphere version that includes ocean dynamics and freezing point depression as in previous exoplanet studies by \citeA{del_genio_habitable_2019} and \citeA{olson_oceanographic_2020} (herein, \textbf{\textit{“Full Physics”}}); 

\item a modified version of the coupled model in which ocean dynamics respond to salinity, but the freezing point of seawater does not respond to salinity (\textbf{\textit{“Fixed Freezing Point”}}); and
 
\item a slab ocean lacking ocean dynamics and heat transport but including freezing point depression (\textbf{\textit{“No Dynamics”}}). 
 \end{enumerate}

In each of our Archean simulations, we decrease solar luminosity to 1106.81 Wm$^{-2}$  \cite{gough_solar_1981}, decrease day length to 17.25 hours \cite{williams_geological_2000}, increase pCH$_4$ to 500 $\mu$bar based on recent model and proxy suggestions \cite{izon_biological_2017, reinhard_oceanic_2020} and decrease total surface pressure to 0.5 bar \cite{som_earths_2016}. We then examine climate sensitivity to varying pCO$_2$ from 10-100 xPIL \cite{hessler_lower_2004, driese_neoarchean_2011} and ocean salinities of 20 g/kg, 35 g/kg (modern), and 50 g/kg. We use the \textit{Full Physics} configuration for all Archean simulations. These experiments are summarized in Table 1.   

We explore salinities of \textpm 15 g/kg relative to present-day seawater because this range is broadly inclusive of existing model and proxy estimates of ancient ocean salinity on Earth \cite{albarede_chemical_2020, catling_archean_2020}, but we note that Archean salinity remains poorly constrained. Our goal is thus not to offer a definitive view of a single moment in Earth’s history; instead, our goal is simply to explore the response of the climate system to changing ocean salinity and to assess the potential significance of these effects in the context of reduced solar luminosity on early Earth. 

We initialize all of our present-day and Archean Earth experiments from a warm, ice-free state. We then diagnose steady state by the achievement of global energy balance $\le$0.2 Wm$^{-2}$ and the convergence of ice coverage and surface temperature. The runtime required to achieve steady state varied between model scenarios, from a few hundred years for snowballs to 1000+ years for warmer climates, but all of the data shown here has been averaged over the last decade of each simulation independent of its total duration.  

\section{Results and Discussion}
\subsection{Present-day Earth}

Increasing ocean salinity results in warming, particularly at high latitudes (Figure 1), and reduced sea ice in all three of our model configurations (Figure 2). Decreasing ocean salinity has the opposite effect, decreasing surface temperature on global average and increasing the extent of sea ice. 

 \begin{figure}
\noindent\includegraphics[width=\textwidth]{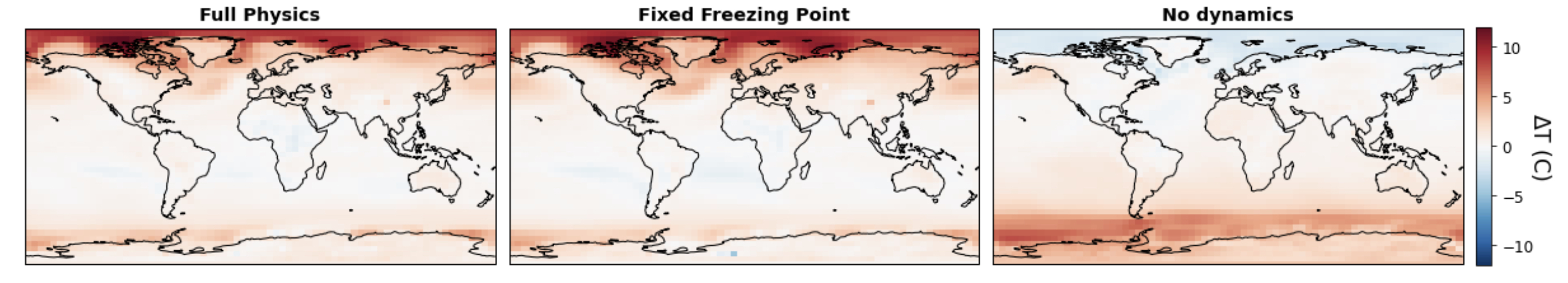}
 \caption{Steady state surface temperature differences for 50 g/kg vs. 20 g/kg salinity scenarios with each model configuration. Reds indicate higher temperatures with higher salinity while blues indicate lower temperatures with higher salinity. All high salinity scenarios are warmer on global average than the low salinity scenarios despite local cooling. }
\end{figure}

The influence of salinity on climate is greatest in our \textit{Full Physics} configuration that includes both density-related dynamic effects and freezing point depression (Figure 2, top row). In our \textit{Full Physics} simulations, increasing salinity from 20 g/kg to 50 g/kg yields warming of up to 11.9 $^{\circ}$C in the annual average at northern high latitudes (0.9 $^{\circ}$C on global average) and a 71\% decrease in global sea ice extent.

Our \textit{Fixed Freezing Point} simulations are similar to the \textit{Full Physics} simulations (Figure 2, middle row). With \textit{Fixed Freezing Point}, increasing salinity from 20 g/kg to 50 g/kg results in warming of up to 11.3 $^{\circ}$C in the annual average at northern high latitudes (0.8 $^{\circ}$C warming on global average) and a 71\% decrease in global sea ice coverage. This similarity to our \textit{Full Physics} simulations suggests that the climate response to changing ocean salinity is primarily due to changing ocean dynamics rather than freezing point depression, which is not included in these experiments. 

 \begin{figure}
\noindent\includegraphics[width=\textwidth]{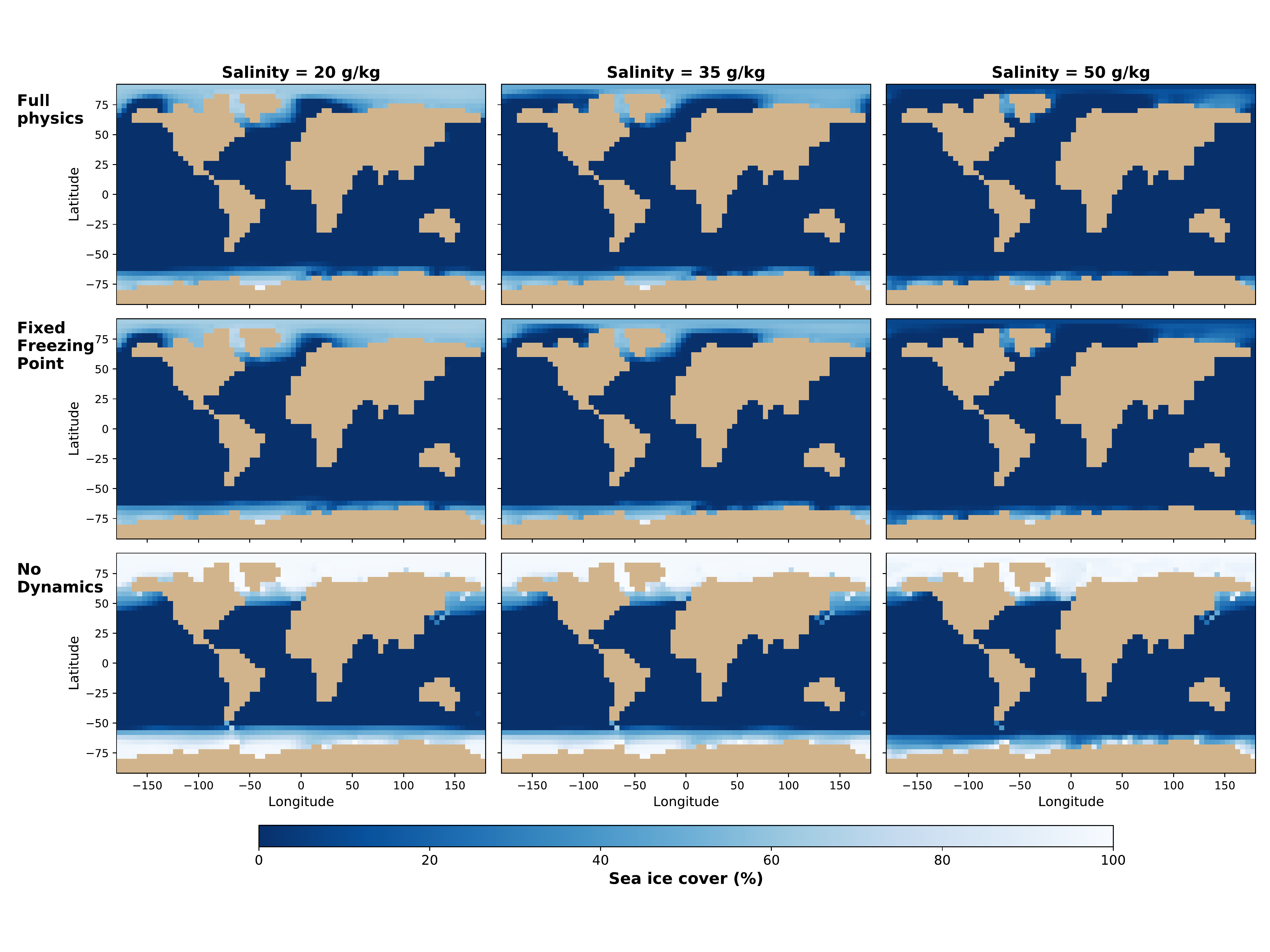}
 \caption{Sea ice coverage for various model configurations and ocean salinities. Rows show the standard model with a fully coupled ocean-atmosphere, sea ice dynamics and freezing point depression (\textit{“Full Physics”}; top); a modified model configuration with ocean and sea ice dynamics but fixed freezing point for all salinities (“\textit{Fixed Freezing Point}”; middle); and a slab ocean configuration lacking ocean and ice dynamics but including freezing point depression (“\textit{No Dynamics}”; bottom). Salinity differs between the columns, increasing from 20 g/kg (left) to 50 g/kg (right). Increasing salinity yields lower ice cover in all scenarios, but the effects are most pronounced in model scenarios that include both dynamical and thermodynamic effects.}
\end{figure}

Our \textit{No Dynamics} simulation with present-day salinity has much greater baseline ice cover than even the low salinity scenarios with our other model configurations (Figure 2, bottom row). This difference arises because the lack of ocean heat transport to high latitudes in these simulations results in colder poles for all salinities. Although increasing (decreasing) ocean salinity does result in warming (cooling) in our \textit{No Dynamics} configuration, the fractional decrease in sea ice (35\%) is small compared to the \textit{Full Physics} and \textit{Fixed Freezing Point} simulations (both 71\%). This relationship again suggests that dynamic effects rather than freezing point depression dominate the response of Earth’s climate to changing salinity. Moreover, the spatial patterns of warming in our \textit{No Dynamics} scenarios differ from our \textit{Full Physics} simulations (Figure 1). 

Warming and ice retreat with increasing salinity are most pronounced at the northern high latitudes in our \textit{Full Physics} and \textit{Fixed Freezing Point} simulations. Conversely, the effects of salinity are greatest in the southern hemisphere in our \textit{No Dynamics} simulations (Figure 1, 2). These results suggest that dynamics have a greater impact on northern hemisphere ice cover. Indeed, the strength of the Atlantic Meridional Overturning Circulation (AMOC) increases with salinity in our experiments (Figure A2), increasing ocean heat transport to northern high latitudes but not to southern high latitudes (Figure A3). Freezing point depression may instead contribute more to sea ice reductions in the southern hemisphere, but our \textit{Fixed Freezing Point} experiments demonstrate that some reduction in sea ice occurs in both hemispheres even without freezing point depression, suggesting that additional dynamical effects are important as well. 

\subsection{Archean Earth} 

Our CO$_2$-salinity parameter space yields several stable Archean climate scenarios including global snowball glaciation, partial glaciation at mid-to-high latitudes similar to present-day Earth, and a nearly ice-free climate that is warmer than present-day Earth (Figure 3). Additionally, many of our simulations achieve a “Water Belt” state that is characterized by low-latitude stabilization of ice and the persistence of a narrow band of equatorial open water \cite{abbot_jormungand_2011, rose_stable_2015}. This Water Belt state is stable against the ice albedo feedback at least in part due to strong overturning circulation extending to the low-latitude ice margin (Figure A4) \cite{rose_stable_2015}. 

The achievement of each Archean climate state is sensitive to both pCO$_2$ and salinity. Increasing either CO$_2$ and/or salinity imparts warming and reduces ice cover. The warmest model scenarios are the highest CO$_2$ and highest salinity scenarios whereas the coldest and iciest simulated climates have the lowest CO$_2$ levels and lowest salinities. Moreover, the CO$_2$ threshold at which the climate system abruptly transitions between these states is itself sensitive to salinity (Figure 4). This sensitivity allows modest changes in salinity to manifest as distinct climate states at fixed pCO$_2$ (Figure 3, 4), and it is likely the consequence of feedbacks between ocean circulation, ice, and surface temperature, which amplify the direct effect of salinity changes. 

 \begin{figure}
\noindent\includegraphics[width=\textwidth]{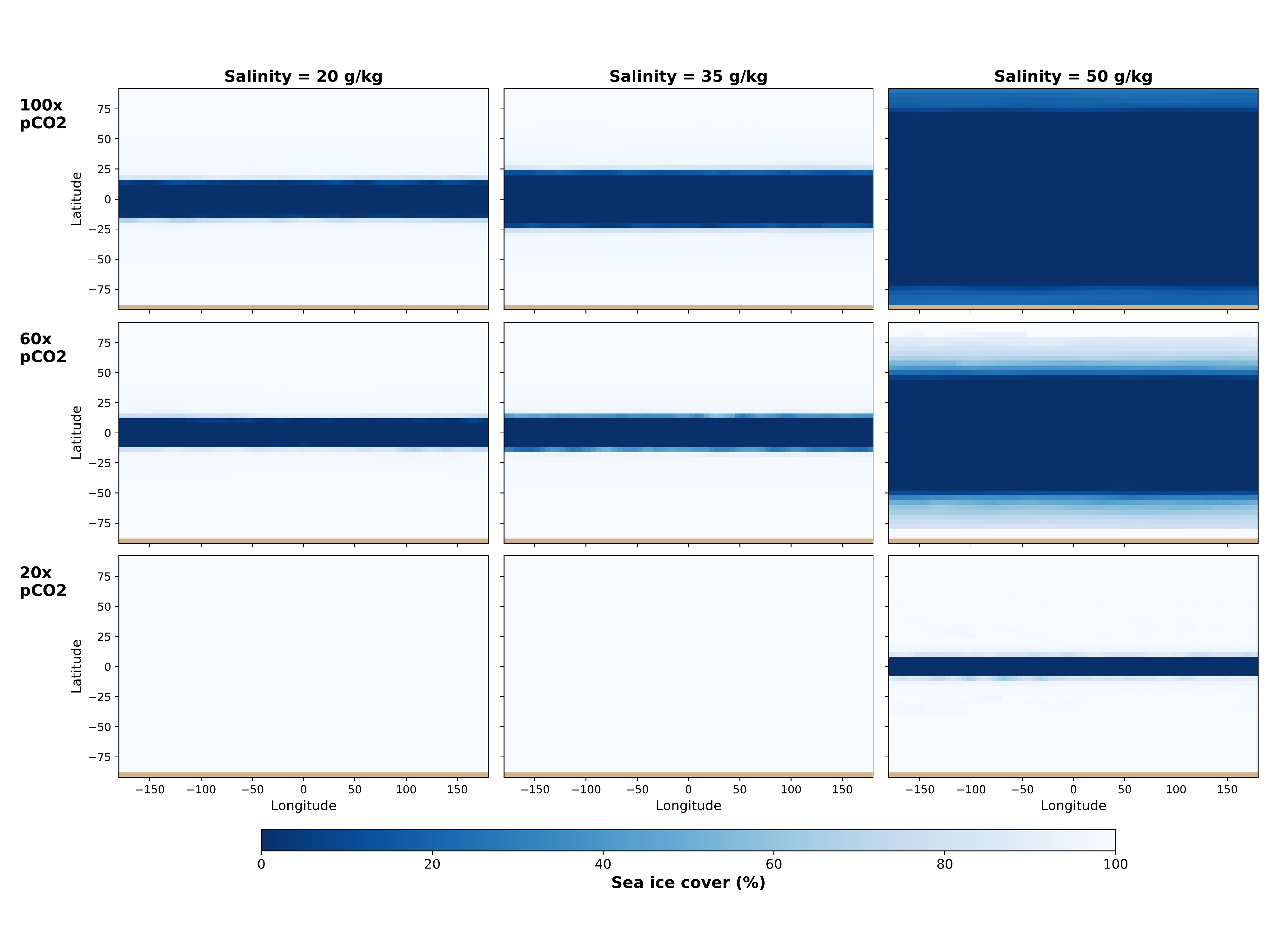}
 \caption{Sea ice concentration for variable salinity and atmospheric CO$_2$. The columns correspond to salinities of 20 g/kg (left), 35 g/kg (middle), and 50 g/kg (right); the rows correspond to CO$_2$ levels of 100 xPIL (top), 60 xPIL (middle), and 20 xPIL (bottom). Dark blue corresponds to open water (0\% ice cover) while white corresponds to 100\% ice cover. Our parameter space yields climate states ranging from snowball glaciation to nearly ice-free conditions. Severely glaciated Water Belt climates that are characterized by a band of equatorial open water are particularly common.  }
\end{figure}

Increasing ocean salinity decreases the CO$_2$ level at which Archaean Earth enters the Water Belt and Snowball states, allowing open water to persist at much lower pCO$_2$ levels and modestly lower global temperatures compared to less salty oceans. An ocean salinity of 50 g/kg allows open water to persist at the equator down to relatively low pCO$_2$ (10-15 xPIL) in our model whereas our modern and low salinity scenarios transition to snowball states at 20-25 xPIL and 35-40 xPIL, respectively (Figure 4). Moreover, the combination of 100 xPIL atmospheric pCO$_2$ and 50 g/kg ocean salinity yields an Archaean climate that is nearly ice free and warmer than present-day Earth. Although high salinity does not eliminate the importance of elevated levels of greenhouse gases relative to present-day levels in our experiments, we note that warmer-than-present-day climates are possible with high salinity despite our somewhat pessimistic assumption of pCO$_2$, pCH$_4$, and surface pressures, all on the lower-end of existing estimates for the Archean \cite{olson_earth:_2018, catling_archean_2020}. 

Higher surface pressure more similar to present-day Earth would enhance equator-to-pole heat transport in the atmosphere \cite{kaspi_atmospheric_2015, komacek_atmospheric_2019}. For scenarios with a high-latitude ice line, this phenomenon is likely to warm the poles and limit ice advance to lower latitudes relative to our scenarios with low surface pressure. However, more efficient atmospheric heat transport away from the equator may destabilize Water Belt climates and promote snowball glaciation with the collapse of the ice line to the equator. Faster rotation also reduces the efficiency of equator-to-pole heat transport and may further contribute to the persistence of low-latitude open water in many of our experiments \cite{spalding_shorter_2019}.

 \begin{figure}
\noindent\includegraphics[width=\textwidth]{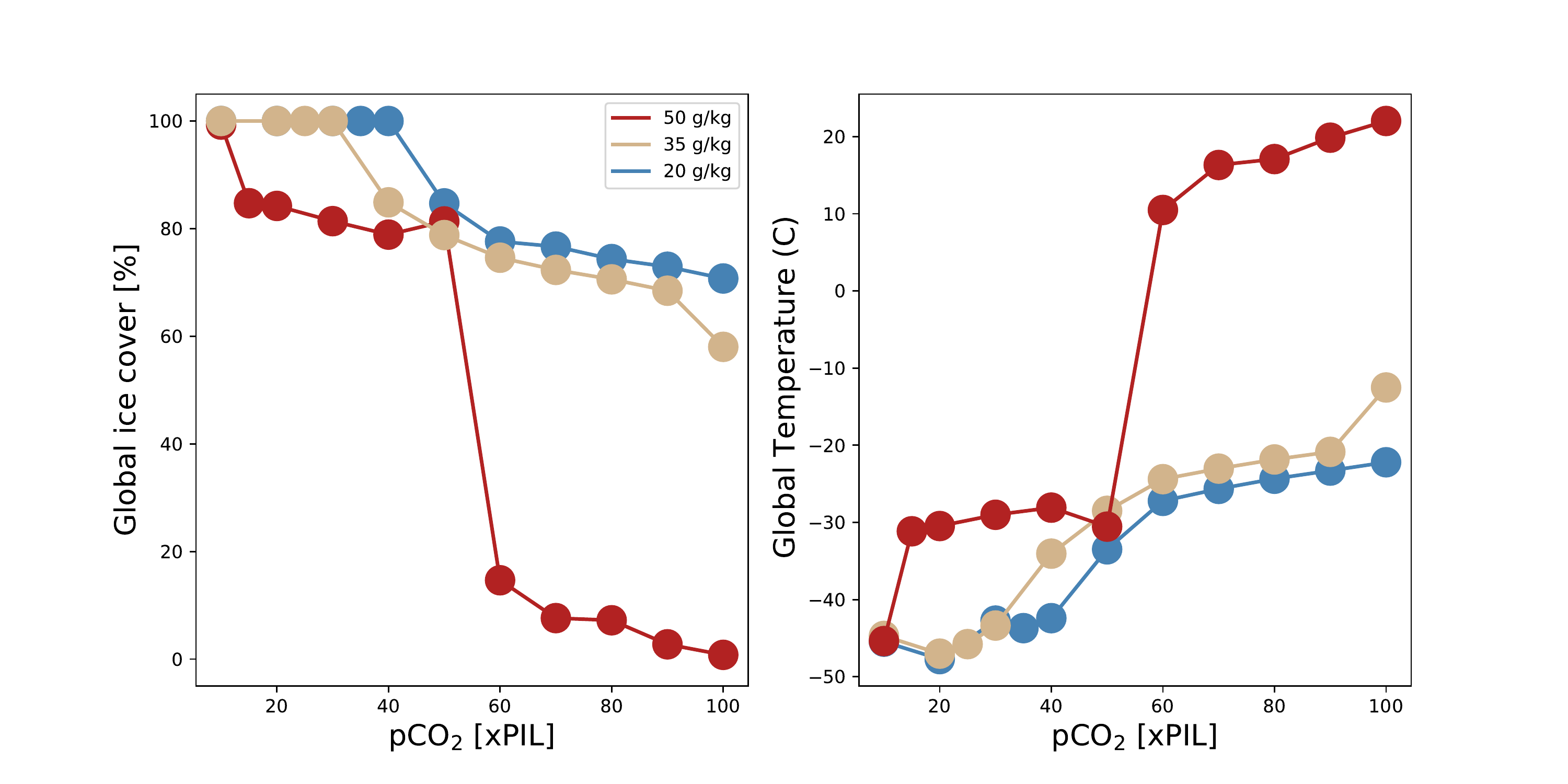}
 \caption{Archean sea ice cover vs. pCO$_2$. The colored lines in each panel correspond to salinities of 20 g/kg (blue), 35 g/kg (green), and 50 g/kg (yellow). Four distinct climate states that are separated by abrupt transitions emerge for the CO$_2$ range considered. The CO$_2$ threshold for the transition between these states is sensitive to ocean salinity, allowing small changes in salinity to profoundly change climate. For example, at 100x PIL CO$_2$, global average surface temperature differs by 45 $^{\circ}$C between the 20 and 50 g/kg salinity scenarios. All simulations were initiated from an ice-free state. }
\end{figure}

\subsection{Implications for Early Earth Habitability}
Ocean salinity is known to be an important consideration for exoplanet habitability, particularly in the outer Habitable Zone \cite{cullum_importance_2016, del_genio_habitable_2019, olson_oceanographic_2020}---not unlike the stellar context of the early Earth around the Faint Young Sun. Our results introduce the tantalizing possibility that a saltier ocean may partially compensate for lower solar luminosity on early Earth. In this case, salt may be an essential ingredient for early Earth habitability. In combination with other warming mechanisms, salt may even contribute to a climate state that was warmer than today. Conversely, our results suggest that a less salty Archean ocean may exacerbate the Faint Young Sun Paradox. Constraining the salinity history of Earth’s ocean is thus an important albeit challenging goal.

\citeA{holland_chemical_1984} estimated that returning known evaporite deposits to the ocean would result in an Archean ocean salinity of ~50 g/kg, similar to the high salinity scenario in our experiments. \citeA{knauth_temperature_2005} went a step further and suggested that if one stirred back basinal brines as well, Archean ocean salinity may have been as high as 70 g/kg. These early estimates are limited by the assumption that (1) essentially all of Earth’s present-day surface Cl inventory was present at Earth’s surface since magma ocean crystallization as the consequence of Cl’s volatility and incompatibility with silicates \cite{kuwahara_fluorine_2019}, and (2) the volume of the ocean has not significantly changed since its formation. Nonetheless, the likelihood of higher atmospheric pCO$_2$ and lower ocean pH (higher [H$^+$]) during the Archean implies a greater abundance of dissolved anions such as Cl$^-$ via charge balance \cite{albarede_chemical_2020}. Indeed, the evaporite record may suggest a secular decline in ocean salinity from ~50 g/kg in the latest Precambrian through the Phanerozoic (Hay et al., 2006), coincident with a long-term decline of atmospheric CO$_2$ and increase in ocean pH \cite{halevy_geologic_2017, krissansen-totton_constraining_2018}. 

\section{Caveats and Opportunities for Future Work}
Salt affects seawater density and ocean dynamics via direct mass effects and through its influence on charge density and ionic interactions with polar water molecules. Salinity is simply input to ROCKE-3D as mass of solute per kg of seawater, with no explicit assumptions about the abundances of specific ions. We thus fully account for the added mass of dissolved ions, but the representation of additional effects (e.g., the dependence of the thermal expansion coefficient on salinity) implicitly assumes a seawater charge balance dominated by monovalent ions as it is today \cite{albarede_chemical_2020}, rather than divalent ions such as Mg$^{2+}$ and SO$_{4}^{2-}$ . 

The magnitude of freezing point depression depends on the chemical identity of the solvent (water), the molality of the solution (mols of solute per kg solvent), and the number of mols of ions resulting from the dissolution of one mol of solute (e.g., 2 mols ions for 1 mol dissolved NaCl)—but it does not depend on the specific chemical identity of the solute. Extending ROCKE-3D’s freezing point calculation to the Archean thus also assumes that Archean seawater had a similar blend of monovalent and divalent ions to the present-day ocean, but this freezing point calculation is insensitive to the likelihood that the Archean ocean had relatively less Mg$^{2+}$ (24 g/mol) due to limited continental weathering and relatively more Fe$^{2+}$ (56 g/mol) due to the scarcity of O$_2$. Although swapping an equivalent molar quantity of Mg$^{2+}$ for relatively heavy Fe$^{2+}$ would not affect the freezing point of seawater, our simulations demonstrate that this change could still affect ocean dynamics through its influence on density.

Ocean salinity has a few additional effects on climate via atmospheric energy balance that are not quantified here. For example, increasing salinity decreases gas solubility and, all else equal, higher salinity should tend to increase greenhouse gas levels in the atmosphere and surface temperatures. We fix atmospheric composition in each of our simulations and thus our experiments do not account for how a ~10\% reduction in gas solubility with increasing salinity between our low and high salinity scenarios would further bolster warming from the oceanographic phenomena we explored. At the same time, increasing salinity reduces the activity of water, decreasing evaporation. This effect results in an observable decrease in evaporation rates over the extremely salty Dead Sea (S=235 g/kg), but for the relatively small range of salinities we explored such effects are expected to be small compared to the changes in evaporation that arise from the differences in surface temperature among our experiments \cite{mor_effect_2018, stanhill_changes_1994}. 

Finally, breaking waves and bubbles inject sea salt into the atmosphere. Sea salt aerosol is primarily modulated by windspeed on present-day Earth \cite{prijith_relationship_2014}. Sea salt aerosol scatters incident shortwave radiation and serves as a cloud condensation nuclei (CCN). These effects are poorly understood on present day Earth and entirely unconstrained on Archean Earth \cite{ma_modelling_2008, pierce_global_2006}. A saltier ocean may result in more sea salt aerosol under equivalent wind and wave conditions, but the likelihood of limited continental shelf area and lower surface pressure (and thus wind stress) on Archean Earth may ultimately decrease the amount of sea salt entering the atmosphere via breaking waves. If CCN abundance increases with salinity, saltier oceans might cause cooling due to greater fractional cloud cover and/or more reflective clouds with smaller droplets \cite{twomey_pollution_1974}---but enhanced cloudiness in this scenario may also weaken the ice-albedo feedback and extend the stability of the Water Belt state to lower insolation and/or CO$_2$ levels \cite{braun_subtropical_2020}. It is thus unclear whether accounting for changes to sea salt aerosol in our model would have a large effect on climate and whether these effects would amplify or offset warming with increasing salinity in our model scenarios. The relationships between ocean salinity, atmospheric water vapor, cloud nucleation, precipitation patterns, and surface temperature on short and long timescales remains an exciting opportunity for future work. 

\section{Conclusions} 
The salient result of our study is that ocean salinity plays an important role in Earth’s climate system. Saltier oceans yield warmer, more equable climates with less sea ice in both our present-day and Archean simulations. Warming with increasing salinity in our simulations is primarily due to changes in ocean dynamics, with a more minor contribution from freezing point depression. The impacts of ocean salinity on climate are also strongly non-linear. While small differences in salinity (\textpm 15 g/kg relative to present day) may yield only modest changes in surface temperature and ice cover in some circumstances, we show that such salinity differences can sometimes result in the transition between distinct climate states. In our Archean scenario with 100x pCO$_2$, present-day ocean salinity resulted in a severely glaciated Water Belt climate with only a narrow strip of open water at the equator—but increasing salinity to 50 g/kg resulted in a warm climate with surface temperatures exceeding 20 $^{\circ}$C on global average and only seasonal ice at the poles. These strongly non-linear effects are difficult to represent in less computationally expensive models lacking ocean dynamics. Key uncertainties for future studies include resolving the details of Earth’s salinity evolution and how ocean salinity affects cloud radiative forcing. With these caveats, we suggest that an Archean ocean that was saltier than today could play a key role in compensating for the Faint Young Sun, perhaps even allowing an Archean climate that was warmer than today. 

%%%%%%%%%%%%%%%%%%%%%%%%%%%%%%%%%%%%%%%%%%%%%%%

\section{Open Research}
ROCKE-3D is available at https://simplex.giss.nasa.gov/gcm/ROCKE-3D/ and described in detail by Way et al. (2017). Data presented herein and essential model input/configuration files are available at:  https://doi.org/10.5281/zenodo.5637502.
\\

%%%%%%%%%%%%%%%%%%%%%%%%%%%%%%%%%%%%%%%%%%%%%%%

\acknowledgments
SLO acknowledges financial support from the NASA Habitable Worlds and NASA Interdisciplinary Consortia for Astrobiology Research (ICAR) programs. This work further benefited from participation in the NASA Nexus for Exoplanet System Science (NExSS) and Network for Ocean Worlds (NOW) Research Coordination Networks. SLO acknowledges the importance of increasing the inclusion of historically marginalized identities in Earth and Planetary Science.

%% ------------------------------------------------------------------------ %%
%% References and Citations

\bibliography{salinity-ms.bib}

\begin{thebibliography}{}

\bibitem [\protect \citeauthoryear {%
Abbot%
, Voigt%
\BCBL {}\ \BBA {} Koll%
}{%
Abbot%
\ \protect \BOthers {.}}{%
{\protect \APACyear {2011}}%
}]{%
abbot_jormungand_2011}
\APACinsertmetastar {%
abbot_jormungand_2011}%
\begin{APACrefauthors}%
Abbot, D\BPBI S.%
, Voigt, A.%
\BCBL {}\ \BBA {} Koll, D.%
\end{APACrefauthors}%
\unskip\
\newblock
\APACrefYearMonthDay{2011}{{\APACmonth{09}}}{}.
\newblock
{\BBOQ}\APACrefatitle {The {Jormungand} global climate state and implications
  for {Neoproterozoic} glaciations} {The {Jormungand} global climate state and
  implications for {Neoproterozoic} glaciations}.{\BBCQ}
\newblock
\APACjournalVolNumPages{Journal of Geophysical Research}{116}{D18}{}.
\newblock
\begin{APACrefDOI} \doi{10.1029/2011JD015927} \end{APACrefDOI}
\PrintBackRefs{\CurrentBib}

\bibitem [\protect \citeauthoryear {%
Albarede%
, Thibon%
, Blichert-Toft%
\BCBL {}\ \BBA {} Tsikos%
}{%
Albarede%
\ \protect \BOthers {.}}{%
{\protect \APACyear {2020}}%
}]{%
albarede_chemical_2020}
\APACinsertmetastar {%
albarede_chemical_2020}%
\begin{APACrefauthors}%
Albarede, F.%
, Thibon, F.%
, Blichert-Toft, J.%
\BCBL {}\ \BBA {} Tsikos, H.%
\end{APACrefauthors}%
\unskip\
\newblock
\APACrefYearMonthDay{2020}{{\APACmonth{08}}}{}.
\newblock
{\BBOQ}\APACrefatitle {Chemical archeoceanography} {Chemical
  archeoceanography}.{\BBCQ}
\newblock
\APACjournalVolNumPages{Chemical Geology}{548}{}{119625}.
\newblock
\begin{APACrefDOI} \doi{10.1016/j.chemgeo.2020.119625} \end{APACrefDOI}
\PrintBackRefs{\CurrentBib}

\bibitem [\protect \citeauthoryear {%
Braun%
, Voigt%
, Hörner%
\BCBL {}\ \BBA {} Pinto%
}{%
Braun%
\ \protect \BOthers {.}}{%
{\protect \APACyear {2020}}%
}]{%
braun_subtropical_2020}
\APACinsertmetastar {%
braun_subtropical_2020}%
\begin{APACrefauthors}%
Braun, C.%
, Voigt, A.%
, Hörner, J.%
\BCBL {}\ \BBA {} Pinto, J\BPBI G.%
\end{APACrefauthors}%
\unskip\
\newblock
\APACrefYearMonthDay{2020}{{\APACmonth{03}}}{}.
\newblock
\APACrefbtitle {Subtropical clouds stabilize near-{Snowball} {Earth} states}
  {Subtropical clouds stabilize near-{Snowball} {Earth} states}\ \APACbVolEdTR
  {}{other}.
\newblock
\begin{APACrefDOI} \doi{10.5194/egusphere-egu2020-10617} \end{APACrefDOI}
\PrintBackRefs{\CurrentBib}

\bibitem [\protect \citeauthoryear {%
Cael%
\ \BBA {} Ferrari%
}{%
Cael%
\ \BBA {} Ferrari%
}{%
{\protect \APACyear {2017}}%
}]{%
cael_oceans_2017}
\APACinsertmetastar {%
cael_oceans_2017}%
\begin{APACrefauthors}%
Cael, B\BPBI B.%
\BCBT {}\ \BBA {} Ferrari, R.%
\end{APACrefauthors}%
\unskip\
\newblock
\APACrefYearMonthDay{2017}{}{}.
\newblock
{\BBOQ}\APACrefatitle {The ocean's saltiness and its overturning} {The ocean's
  saltiness and its overturning}.{\BBCQ}
\newblock
\APACjournalVolNumPages{Geophysical Research Letters}{44}{4}{1886--1891}.
\newblock
\begin{APACrefDOI} \doi{10.1002/2016GL072223} \end{APACrefDOI}
\PrintBackRefs{\CurrentBib}

\bibitem [\protect \citeauthoryear {%
Catling%
\ \BBA {} Zahnle%
}{%
Catling%
\ \BBA {} Zahnle%
}{%
{\protect \APACyear {2020}}%
}]{%
catling_archean_2020}
\APACinsertmetastar {%
catling_archean_2020}%
\begin{APACrefauthors}%
Catling, D\BPBI C.%
\BCBT {}\ \BBA {} Zahnle, K\BPBI J.%
\end{APACrefauthors}%
\unskip\
\newblock
\APACrefYearMonthDay{2020}{{\APACmonth{02}}}{}.
\newblock
{\BBOQ}\APACrefatitle {The {Archean} atmosphere} {The {Archean}
  atmosphere}.{\BBCQ}
\newblock
\APACjournalVolNumPages{Science Advances}{6}{9}{eaax1420}.
\newblock
\begin{APACrefDOI} \doi{10.1126/sciadv.aax1420} \end{APACrefDOI}
\PrintBackRefs{\CurrentBib}

\bibitem [\protect \citeauthoryear {%
Catling%
, Zahnle%
\BCBL {}\ \BBA {} McKay%
}{%
Catling%
\ \protect \BOthers {.}}{%
{\protect \APACyear {2001}}%
}]{%
catling_biogenic_2001}
\APACinsertmetastar {%
catling_biogenic_2001}%
\begin{APACrefauthors}%
Catling, D\BPBI C.%
, Zahnle, K\BPBI J.%
\BCBL {}\ \BBA {} McKay, C\BPBI P.%
\end{APACrefauthors}%
\unskip\
\newblock
\APACrefYearMonthDay{2001}{}{}.
\newblock
{\BBOQ}\APACrefatitle {Biogenic methane, hydrogen escape, and the irreversible
  oxidation of early {Earth}} {Biogenic methane, hydrogen escape, and the
  irreversible oxidation of early {Earth}}.{\BBCQ}
\newblock
\APACjournalVolNumPages{Science}{293}{5531}{839--843}.
\newblock
\begin{APACrefDOI} \doi{10.1126/science.1061976} \end{APACrefDOI}
\PrintBackRefs{\CurrentBib}

\bibitem [\protect \citeauthoryear {%
Cullum%
, Stevens%
\BCBL {}\ \BBA {} Joshi%
}{%
Cullum%
\ \protect \BOthers {.}}{%
{\protect \APACyear {2016}}%
}]{%
cullum_importance_2016}
\APACinsertmetastar {%
cullum_importance_2016}%
\begin{APACrefauthors}%
Cullum, J.%
, Stevens, D\BPBI P.%
\BCBL {}\ \BBA {} Joshi, M\BPBI M.%
\end{APACrefauthors}%
\unskip\
\newblock
\APACrefYearMonthDay{2016}{{\APACmonth{04}}}{}.
\newblock
{\BBOQ}\APACrefatitle {Importance of ocean salinity for climate and
  habitability} {Importance of ocean salinity for climate and
  habitability}.{\BBCQ}
\newblock
\APACjournalVolNumPages{Proceedings of the National Academy of
  Sciences}{113}{16}{4278--4283}.
\newblock
\begin{APACrefDOI} \doi{10.1073/pnas.1522034113} \end{APACrefDOI}
\PrintBackRefs{\CurrentBib}

\bibitem [\protect \citeauthoryear {%
Del~Genio%
\ \protect \BOthers {.}}{%
Del~Genio%
\ \protect \BOthers {.}}{%
{\protect \APACyear {2019}}%
}]{%
del_genio_habitable_2019}
\APACinsertmetastar {%
del_genio_habitable_2019}%
\begin{APACrefauthors}%
Del~Genio, A\BPBI D.%
, Way, M\BPBI J.%
, Amundsen, D\BPBI S.%
, Aleinov, I.%
, Kelley, M.%
, Kiang, N\BPBI Y.%
\BCBL {}\ \BBA {} Clune, T\BPBI L.%
\end{APACrefauthors}%
\unskip\
\newblock
\APACrefYearMonthDay{2019}{}{}.
\newblock
{\BBOQ}\APACrefatitle {Habitable {Climate} {Scenarios} for {Proxima} {Centauri}
  b with a {Dynamic} {Ocean}} {Habitable {Climate} {Scenarios} for {Proxima}
  {Centauri} b with a {Dynamic} {Ocean}}.{\BBCQ}
\newblock
\APACjournalVolNumPages{Astrobiology}{19}{}{}.
\PrintBackRefs{\CurrentBib}

\bibitem [\protect \citeauthoryear {%
Driese%
\ \protect \BOthers {.}}{%
Driese%
\ \protect \BOthers {.}}{%
{\protect \APACyear {2011}}%
}]{%
driese_neoarchean_2011}
\APACinsertmetastar {%
driese_neoarchean_2011}%
\begin{APACrefauthors}%
Driese, S\BPBI G.%
, Jirsa, M\BPBI A.%
, Ren, M.%
, Brantley, S\BPBI L.%
, Sheldon, N\BPBI D.%
, Parker, D.%
\BCBL {}\ \BBA {} Schmitz, M.%
\end{APACrefauthors}%
\unskip\
\newblock
\APACrefYearMonthDay{2011}{}{}.
\newblock
{\BBOQ}\APACrefatitle {Neoarchean paleoweathering of tonalite and metabasalt:
  {Implications} for reconstructions of 2.{69Ga} early terrestrial ecosystems
  and paleoatmospheric chemistry} {Neoarchean paleoweathering of tonalite and
  metabasalt: {Implications} for reconstructions of 2.{69Ga} early terrestrial
  ecosystems and paleoatmospheric chemistry}.{\BBCQ}
\newblock
\APACjournalVolNumPages{Precambrian Research}{189}{1}{1--17}.
\newblock
\begin{APACrefDOI} \doi{10.1016/j.precamres.2011.04.003} \end{APACrefDOI}
\PrintBackRefs{\CurrentBib}

\bibitem [\protect \citeauthoryear {%
Emerson%
\ \BBA {} Hedges%
}{%
Emerson%
\ \BBA {} Hedges%
}{%
{\protect \APACyear {2008}}%
}]{%
emerson_chemical_2008}
\APACinsertmetastar {%
emerson_chemical_2008}%
\begin{APACrefauthors}%
Emerson, S.%
\BCBT {}\ \BBA {} Hedges, J.%
\end{APACrefauthors}%
\unskip\
\newblock
\APACrefYear{2008}.
\newblock
\APACrefbtitle {Chemical {Oceanography} and the {Marine} {Carbon} {Cycle}}
  {Chemical {Oceanography} and the {Marine} {Carbon} {Cycle}}.
\newblock
\APACaddressPublisher{}{Cambridge University Press}.
\PrintBackRefs{\CurrentBib}

\bibitem [\protect \citeauthoryear {%
Feulner%
}{%
Feulner%
}{%
{\protect \APACyear {2012}}%
}]{%
feulner_faint_2012}
\APACinsertmetastar {%
feulner_faint_2012}%
\begin{APACrefauthors}%
Feulner, G.%
\end{APACrefauthors}%
\unskip\
\newblock
\APACrefYearMonthDay{2012}{{\APACmonth{06}}}{}.
\newblock
{\BBOQ}\APACrefatitle {The faint young {Sun} problem} {The faint young {Sun}
  problem}.{\BBCQ}
\newblock
\APACjournalVolNumPages{Reviews of Geophysics}{50}{2}{}.
\newblock
\begin{APACrefDOI} \doi{10.1029/2011RG000375} \end{APACrefDOI}
\PrintBackRefs{\CurrentBib}

\bibitem [\protect \citeauthoryear {%
Fofonoff%
\ \BBA {} Millard%
}{%
Fofonoff%
\ \BBA {} Millard%
}{%
{\protect \APACyear {1983}}%
}]{%
fofonoff_algorithms_1983}
\APACinsertmetastar {%
fofonoff_algorithms_1983}%
\begin{APACrefauthors}%
Fofonoff, N.%
\BCBT {}\ \BBA {} Millard, R.%
\end{APACrefauthors}%
\unskip\
\newblock
\APACrefYearMonthDay{1983}{}{}.
\newblock
{\BBOQ}\APACrefatitle {Algorithms for computation of fundamental properites of
  seawater} {Algorithms for computation of fundamental properites of
  seawater}.{\BBCQ}
\newblock
\APACjournalVolNumPages{UNESCO Technical Papers in Marine Science}{44}{}{}.
\PrintBackRefs{\CurrentBib}

\bibitem [\protect \citeauthoryear {%
Gough%
}{%
Gough%
}{%
{\protect \APACyear {1981}}%
}]{%
gough_solar_1981}
\APACinsertmetastar {%
gough_solar_1981}%
\begin{APACrefauthors}%
Gough, D\BPBI O.%
\end{APACrefauthors}%
\unskip\
\newblock
\APACrefYearMonthDay{1981}{}{}.
\newblock
{\BBOQ}\APACrefatitle {Solar interior structure and luminosity variations}
  {Solar interior structure and luminosity variations}.{\BBCQ}
\newblock
\APACjournalVolNumPages{Sol. Phys.}{74}{}{21--34}.
\newblock
\begin{APACrefDOI} \doi{10.1007/BF00151270} \end{APACrefDOI}
\PrintBackRefs{\CurrentBib}

\bibitem [\protect \citeauthoryear {%
Halevy%
\ \BBA {} Bachan%
}{%
Halevy%
\ \BBA {} Bachan%
}{%
{\protect \APACyear {2017}}%
}]{%
halevy_geologic_2017}
\APACinsertmetastar {%
halevy_geologic_2017}%
\begin{APACrefauthors}%
Halevy, I.%
\BCBT {}\ \BBA {} Bachan, A.%
\end{APACrefauthors}%
\unskip\
\newblock
\APACrefYearMonthDay{2017}{{\APACmonth{03}}}{}.
\newblock
{\BBOQ}\APACrefatitle {The geologic history of seawater {pH}} {The geologic
  history of seawater {pH}}.{\BBCQ}
\newblock
\APACjournalVolNumPages{Science}{355}{6329}{1069--1071}.
\newblock
\begin{APACrefDOI} \doi{10.1126/science.aal4151} \end{APACrefDOI}
\PrintBackRefs{\CurrentBib}

\bibitem [\protect \citeauthoryear {%
Hay%
\ \protect \BOthers {.}}{%
Hay%
\ \protect \BOthers {.}}{%
{\protect \APACyear {2006}}%
}]{%
hay_evaporites_2006}
\APACinsertmetastar {%
hay_evaporites_2006}%
\begin{APACrefauthors}%
Hay, W\BPBI W.%
, Migdisov, A.%
, Balukhovsky, A\BPBI N.%
, Wold, C\BPBI N.%
, Flögel, S.%
\BCBL {}\ \BBA {} Söding, E.%
\end{APACrefauthors}%
\unskip\
\newblock
\APACrefYearMonthDay{2006}{{\APACmonth{10}}}{}.
\newblock
{\BBOQ}\APACrefatitle {Evaporites and the salinity of the ocean during the
  {Phanerozoic}: {Implications} for climate, ocean circulation and life}
  {Evaporites and the salinity of the ocean during the {Phanerozoic}:
  {Implications} for climate, ocean circulation and life}.{\BBCQ}
\newblock
\APACjournalVolNumPages{Palaeogeography, Palaeoclimatology,
  Palaeoecology}{240}{1-2}{3--46}.
\newblock
\begin{APACrefDOI} \doi{10.1016/j.palaeo.2006.03.044} \end{APACrefDOI}
\PrintBackRefs{\CurrentBib}

\bibitem [\protect \citeauthoryear {%
Hessler%
, Lowe%
, Jones%
\BCBL {}\ \BBA {} Bird%
}{%
Hessler%
\ \protect \BOthers {.}}{%
{\protect \APACyear {2004}}%
}]{%
hessler_lower_2004}
\APACinsertmetastar {%
hessler_lower_2004}%
\begin{APACrefauthors}%
Hessler, A\BPBI M.%
, Lowe, D\BPBI R.%
, Jones, R\BPBI L.%
\BCBL {}\ \BBA {} Bird, D\BPBI K.%
\end{APACrefauthors}%
\unskip\
\newblock
\APACrefYearMonthDay{2004}{}{}.
\newblock
{\BBOQ}\APACrefatitle {A lower limit for atmospheric carbon dioxide levels 3.2
  billion years ago} {A lower limit for atmospheric carbon dioxide levels 3.2
  billion years ago}.{\BBCQ}
\newblock
\APACjournalVolNumPages{Nature}{428}{}{736--738}.
\newblock
\begin{APACrefDOI} \doi{10.1038/nature02471} \end{APACrefDOI}
\PrintBackRefs{\CurrentBib}

\bibitem [\protect \citeauthoryear {%
Holland%
}{%
Holland%
}{%
{\protect \APACyear {1984}}%
}]{%
holland_chemical_1984}
\APACinsertmetastar {%
holland_chemical_1984}%
\begin{APACrefauthors}%
Holland, H\BPBI D.%
\end{APACrefauthors}%
\unskip\
\newblock
\APACrefYear{1984}.
\newblock
\APACrefbtitle {The chemical evolution of the atmosphere and oceans} {The
  chemical evolution of the atmosphere and oceans}.
\newblock
\APACaddressPublisher{Princeton, N.J}{Princeton University Press}.
\PrintBackRefs{\CurrentBib}

\bibitem [\protect \citeauthoryear {%
Isson%
\ \BBA {} Planavsky%
}{%
Isson%
\ \BBA {} Planavsky%
}{%
{\protect \APACyear {2018}}%
}]{%
isson_reverse_2018}
\APACinsertmetastar {%
isson_reverse_2018}%
\begin{APACrefauthors}%
Isson, T\BPBI T.%
\BCBT {}\ \BBA {} Planavsky, N\BPBI J.%
\end{APACrefauthors}%
\unskip\
\newblock
\APACrefYearMonthDay{2018}{{\APACmonth{08}}}{}.
\newblock
{\BBOQ}\APACrefatitle {Reverse weathering as a long-term stabilizer of marine
  {pH} and planetary climate} {Reverse weathering as a long-term stabilizer of
  marine {pH} and planetary climate}.{\BBCQ}
\newblock
\APACjournalVolNumPages{Nature}{560}{7719}{471--475}.
\newblock
\begin{APACrefDOI} \doi{10.1038/s41586-018-0408-4} \end{APACrefDOI}
\PrintBackRefs{\CurrentBib}

\bibitem [\protect \citeauthoryear {%
Izon%
\ \protect \BOthers {.}}{%
Izon%
\ \protect \BOthers {.}}{%
{\protect \APACyear {2017}}%
}]{%
izon_biological_2017}
\APACinsertmetastar {%
izon_biological_2017}%
\begin{APACrefauthors}%
Izon, G.%
, Zerkle, A\BPBI L.%
, Williford, K\BPBI H.%
, Farquhar, J.%
, Poulton, S\BPBI W.%
\BCBL {}\ \BBA {} Claire, M\BPBI W.%
\end{APACrefauthors}%
\unskip\
\newblock
\APACrefYearMonthDay{2017}{}{}.
\newblock
{\BBOQ}\APACrefatitle {Biological regulation of atmospheric chemistry en route
  to planetary oxygenation} {Biological regulation of atmospheric chemistry en
  route to planetary oxygenation}.{\BBCQ}
\newblock
\APACjournalVolNumPages{Proceedings of the National Academy of
  Sciences}{114}{13}{E2571--E2579}.
\newblock
\begin{APACrefDOI} \doi{10.1073/pnas.1618798114} \end{APACrefDOI}
\PrintBackRefs{\CurrentBib}

\bibitem [\protect \citeauthoryear {%
Kaspi%
\ \BBA {} Showman%
}{%
Kaspi%
\ \BBA {} Showman%
}{%
{\protect \APACyear {2015}}%
}]{%
kaspi_atmospheric_2015}
\APACinsertmetastar {%
kaspi_atmospheric_2015}%
\begin{APACrefauthors}%
Kaspi, Y.%
\BCBT {}\ \BBA {} Showman, A\BPBI P.%
\end{APACrefauthors}%
\unskip\
\newblock
\APACrefYearMonthDay{2015}{{\APACmonth{05}}}{}.
\newblock
{\BBOQ}\APACrefatitle {{ATMOSPHERIC} {DYNAMICS} {OF} {TERRESTRIAL} {EXOPLANETS}
  {OVER} {A} {WIDE} {RANGE} {OF} {ORBITAL} {AND} {ATMOSPHERIC} {PARAMETERS}}
  {{ATMOSPHERIC} {DYNAMICS} {OF} {TERRESTRIAL} {EXOPLANETS} {OVER} {A} {WIDE}
  {RANGE} {OF} {ORBITAL} {AND} {ATMOSPHERIC} {PARAMETERS}}.{\BBCQ}
\newblock
\APACjournalVolNumPages{The Astrophysical Journal}{804}{1}{60}.
\newblock
\begin{APACrefDOI} \doi{10.1088/0004-637X/804/1/60} \end{APACrefDOI}
\PrintBackRefs{\CurrentBib}

\bibitem [\protect \citeauthoryear {%
Kasting%
}{%
Kasting%
}{%
{\protect \APACyear {2005}}%
}]{%
kasting_methane_2005}
\APACinsertmetastar {%
kasting_methane_2005}%
\begin{APACrefauthors}%
Kasting, J\BPBI F.%
\end{APACrefauthors}%
\unskip\
\newblock
\APACrefYearMonthDay{2005}{}{}.
\newblock
{\BBOQ}\APACrefatitle {Methane and climate during the {Precambrian} era}
  {Methane and climate during the {Precambrian} era}.{\BBCQ}
\newblock
\APACjournalVolNumPages{Precambrian Research}{137}{3-4}{119--129}.
\newblock
\begin{APACrefDOI} \doi{10.1016/j.precamres.2005.03.002} \end{APACrefDOI}
\PrintBackRefs{\CurrentBib}

\bibitem [\protect \citeauthoryear {%
Knauth%
}{%
Knauth%
}{%
{\protect \APACyear {2005}}%
}]{%
knauth_temperature_2005}
\APACinsertmetastar {%
knauth_temperature_2005}%
\begin{APACrefauthors}%
Knauth, L\BPBI P.%
\end{APACrefauthors}%
\unskip\
\newblock
\APACrefYearMonthDay{2005}{{\APACmonth{04}}}{}.
\newblock
{\BBOQ}\APACrefatitle {Temperature and salinity history of the {Precambrian}
  ocean: implications for the course of microbial evolution} {Temperature and
  salinity history of the {Precambrian} ocean: implications for the course of
  microbial evolution}.{\BBCQ}
\newblock
\APACjournalVolNumPages{Palaeogeography, Palaeoclimatology,
  Palaeoecology}{219}{1-2}{53--69}.
\newblock
\begin{APACrefDOI} \doi{10.1016/j.palaeo.2004.10.014} \end{APACrefDOI}
\PrintBackRefs{\CurrentBib}

\bibitem [\protect \citeauthoryear {%
Komacek%
\ \BBA {} Abbot%
}{%
Komacek%
\ \BBA {} Abbot%
}{%
{\protect \APACyear {2019}}%
}]{%
komacek_atmospheric_2019}
\APACinsertmetastar {%
komacek_atmospheric_2019}%
\begin{APACrefauthors}%
Komacek, T\BPBI D.%
\BCBT {}\ \BBA {} Abbot, D\BPBI S.%
\end{APACrefauthors}%
\unskip\
\newblock
\APACrefYearMonthDay{2019}{{\APACmonth{02}}}{}.
\newblock
{\BBOQ}\APACrefatitle {The {Atmospheric} {Circulation} and {Climate} of
  {Terrestrial} {Planets} {Orbiting} {Sun}-like and {M} {Dwarf} {Stars} over a
  {Broad} {Range} of {Planetary} {Parameters}} {The {Atmospheric} {Circulation}
  and {Climate} of {Terrestrial} {Planets} {Orbiting} {Sun}-like and {M}
  {Dwarf} {Stars} over a {Broad} {Range} of {Planetary} {Parameters}}.{\BBCQ}
\newblock
\APACjournalVolNumPages{The Astrophysical Journal}{871}{2}{245}.
\newblock
\begin{APACrefDOI} \doi{10.3847/1538-4357/aafb33} \end{APACrefDOI}
\PrintBackRefs{\CurrentBib}

\bibitem [\protect \citeauthoryear {%
Korenaga%
, Planavsky%
\BCBL {}\ \BBA {} Evans%
}{%
Korenaga%
\ \protect \BOthers {.}}{%
{\protect \APACyear {2017}}%
}]{%
korenaga_global_2017}
\APACinsertmetastar {%
korenaga_global_2017}%
\begin{APACrefauthors}%
Korenaga, J.%
, Planavsky, N\BPBI J.%
\BCBL {}\ \BBA {} Evans, D\BPBI A\BPBI D.%
\end{APACrefauthors}%
\unskip\
\newblock
\APACrefYearMonthDay{2017}{{\APACmonth{05}}}{}.
\newblock
{\BBOQ}\APACrefatitle {Global water cycle and the coevolution of the
  {Earth}’s interior and surface environment} {Global water cycle and the
  coevolution of the {Earth}’s interior and surface environment}.{\BBCQ}
\newblock
\APACjournalVolNumPages{Philosophical Transactions of the Royal Society A:
  Mathematical, Physical and Engineering Sciences}{375}{2094}{20150393}.
\newblock
\begin{APACrefDOI} \doi{10.1098/rsta.2015.0393} \end{APACrefDOI}
\PrintBackRefs{\CurrentBib}

\bibitem [\protect \citeauthoryear {%
Krissansen-Totton%
, Arney%
\BCBL {}\ \BBA {} Catling%
}{%
Krissansen-Totton%
\ \protect \BOthers {.}}{%
{\protect \APACyear {2018}}%
}]{%
krissansen-totton_constraining_2018}
\APACinsertmetastar {%
krissansen-totton_constraining_2018}%
\begin{APACrefauthors}%
Krissansen-Totton, J.%
, Arney, G\BPBI N.%
\BCBL {}\ \BBA {} Catling, D\BPBI C.%
\end{APACrefauthors}%
\unskip\
\newblock
\APACrefYearMonthDay{2018}{{\APACmonth{04}}}{}.
\newblock
{\BBOQ}\APACrefatitle {Constraining the climate and ocean {pH} of the early
  {Earth} with a geological carbon cycle model} {Constraining the climate and
  ocean {pH} of the early {Earth} with a geological carbon cycle model}.{\BBCQ}
\newblock
\APACjournalVolNumPages{Proceedings of the National Academy of
  Sciences}{115}{16}{4105--4110}.
\newblock
\begin{APACrefDOI} \doi{10.1073/pnas.1721296115} \end{APACrefDOI}
\PrintBackRefs{\CurrentBib}

\bibitem [\protect \citeauthoryear {%
Kuwahara%
\ \protect \BOthers {.}}{%
Kuwahara%
\ \protect \BOthers {.}}{%
{\protect \APACyear {2019}}%
}]{%
kuwahara_fluorine_2019}
\APACinsertmetastar {%
kuwahara_fluorine_2019}%
\begin{APACrefauthors}%
Kuwahara, H.%
, Kagoshima, T.%
, Nakada, R.%
, Ogawa, N.%
, Yamaguchi, A.%
, Sano, Y.%
\BCBL {}\ \BBA {} Irifune, T.%
\end{APACrefauthors}%
\unskip\
\newblock
\APACrefYearMonthDay{2019}{{\APACmonth{08}}}{}.
\newblock
{\BBOQ}\APACrefatitle {Fluorine and chlorine fractionation during magma ocean
  crystallization: {Constraints} on the origin of the non-chondritic {F}/{Cl}
  ratio of the {Earth}} {Fluorine and chlorine fractionation during magma ocean
  crystallization: {Constraints} on the origin of the non-chondritic {F}/{Cl}
  ratio of the {Earth}}.{\BBCQ}
\newblock
\APACjournalVolNumPages{Earth and Planetary Science Letters}{520}{}{241--249}.
\newblock
\begin{APACrefDOI} \doi{10.1016/j.epsl.2019.05.041} \end{APACrefDOI}
\PrintBackRefs{\CurrentBib}

\bibitem [\protect \citeauthoryear {%
Lyons%
, Reinhard%
\BCBL {}\ \BBA {} Planavsky%
}{%
Lyons%
\ \protect \BOthers {.}}{%
{\protect \APACyear {2014}}%
}]{%
lyons_rise_2014}
\APACinsertmetastar {%
lyons_rise_2014}%
\begin{APACrefauthors}%
Lyons, T\BPBI W.%
, Reinhard, C\BPBI T.%
\BCBL {}\ \BBA {} Planavsky, N\BPBI J.%
\end{APACrefauthors}%
\unskip\
\newblock
\APACrefYearMonthDay{2014}{}{}.
\newblock
{\BBOQ}\APACrefatitle {The rise of oxygen in {Earth}'s early ocean and
  atmosphere} {The rise of oxygen in {Earth}'s early ocean and
  atmosphere}.{\BBCQ}
\newblock
\APACjournalVolNumPages{Nature}{506}{7488}{307--315}.
\newblock
\begin{APACrefDOI} \doi{10.1038/nature13068} \end{APACrefDOI}
\PrintBackRefs{\CurrentBib}

\bibitem [\protect \citeauthoryear {%
Ma%
, von Salzen%
\BCBL {}\ \BBA {} Li%
}{%
Ma%
\ \protect \BOthers {.}}{%
{\protect \APACyear {2008}}%
}]{%
ma_modelling_2008}
\APACinsertmetastar {%
ma_modelling_2008}%
\begin{APACrefauthors}%
Ma, X.%
, von Salzen, K.%
\BCBL {}\ \BBA {} Li, J.%
\end{APACrefauthors}%
\unskip\
\newblock
\APACrefYearMonthDay{2008}{{\APACmonth{03}}}{}.
\newblock
{\BBOQ}\APACrefatitle {Modelling sea salt aerosol and its direct and indirect
  effects on climate} {Modelling sea salt aerosol and its direct and indirect
  effects on climate}.{\BBCQ}
\newblock
\APACjournalVolNumPages{Atmospheric Chemistry and Physics}{8}{5}{1311--1327}.
\newblock
\begin{APACrefDOI} \doi{10.5194/acp-8-1311-2008} \end{APACrefDOI}
\PrintBackRefs{\CurrentBib}

\bibitem [\protect \citeauthoryear {%
Mor%
, Assouline%
, Tanny%
, Lensky%
\BCBL {}\ \BBA {} Lensky%
}{%
Mor%
\ \protect \BOthers {.}}{%
{\protect \APACyear {2018}}%
}]{%
mor_effect_2018}
\APACinsertmetastar {%
mor_effect_2018}%
\begin{APACrefauthors}%
Mor, Z.%
, Assouline, S.%
, Tanny, J.%
, Lensky, I\BPBI M.%
\BCBL {}\ \BBA {} Lensky, N\BPBI G.%
\end{APACrefauthors}%
\unskip\
\newblock
\APACrefYearMonthDay{2018}{{\APACmonth{03}}}{}.
\newblock
{\BBOQ}\APACrefatitle {Effect of {Water} {Surface} {Salinity} on {Evaporation}:
  {The} {Case} of a {Diluted} {Buoyant} {Plume} {Over} the {Dead} {Sea}}
  {Effect of {Water} {Surface} {Salinity} on {Evaporation}: {The} {Case} of a
  {Diluted} {Buoyant} {Plume} {Over} the {Dead} {Sea}}.{\BBCQ}
\newblock
\APACjournalVolNumPages{Water Resources Research}{54}{3}{1460--1475}.
\newblock
\begin{APACrefDOI} \doi{10.1002/2017WR021995} \end{APACrefDOI}
\PrintBackRefs{\CurrentBib}

\bibitem [\protect \citeauthoryear {%
Olson%
, Jansen%
\BCBL {}\ \BBA {} Abbot%
}{%
Olson%
\ \protect \BOthers {.}}{%
{\protect \APACyear {2020}}%
}]{%
olson_oceanographic_2020}
\APACinsertmetastar {%
olson_oceanographic_2020}%
\begin{APACrefauthors}%
Olson, S\BPBI L.%
, Jansen, M.%
\BCBL {}\ \BBA {} Abbot, D\BPBI S.%
\end{APACrefauthors}%
\unskip\
\newblock
\APACrefYearMonthDay{2020}{{\APACmonth{05}}}{}.
\newblock
{\BBOQ}\APACrefatitle {Oceanographic {Considerations} for {Exoplanet} {Life}
  {Detection}} {Oceanographic {Considerations} for {Exoplanet} {Life}
  {Detection}}.{\BBCQ}
\newblock
\APACjournalVolNumPages{The Astrophysical Journal}{895}{1}{19}.
\newblock
\begin{APACrefDOI} \doi{10.3847/1538-4357/ab88c9} \end{APACrefDOI}
\PrintBackRefs{\CurrentBib}

\bibitem [\protect \citeauthoryear {%
Olson%
, Schwieterman%
, Reinhard%
\BCBL {}\ \BBA {} Lyons%
}{%
Olson%
\ \protect \BOthers {.}}{%
{\protect \APACyear {2018}}%
}]{%
olson_earth:_2018}
\APACinsertmetastar {%
olson_earth:_2018}%
\begin{APACrefauthors}%
Olson, S\BPBI L.%
, Schwieterman, E\BPBI W.%
, Reinhard, C\BPBI T.%
\BCBL {}\ \BBA {} Lyons, T\BPBI W.%
\end{APACrefauthors}%
\unskip\
\newblock
\APACrefYearMonthDay{2018}{}{}.
\newblock
{\BBOQ}\APACrefatitle {Earth: {Atmospheric} {Evolution} of a {Habitable}
  {Planet}} {Earth: {Atmospheric} {Evolution} of a {Habitable}
  {Planet}}.{\BBCQ}
\newblock
\BIn{} H\BPBI J.~Deeg\ \BBA {} J\BPBI A.~Belmonte\ (\BEDS), \APACrefbtitle
  {Handbook of {Exoplanets}} {Handbook of {Exoplanets}}\ (\BPGS\ 1--37).
\newblock
\APACaddressPublisher{Cham}{Springer International Publishing}.
\PrintBackRefs{\CurrentBib}

\bibitem [\protect \citeauthoryear {%
Pavlov%
, Hurtgen%
, Kasting%
\BCBL {}\ \BBA {} Arthur%
}{%
Pavlov%
\ \protect \BOthers {.}}{%
{\protect \APACyear {2003}}%
}]{%
pavlov_methane-rich_2003}
\APACinsertmetastar {%
pavlov_methane-rich_2003}%
\begin{APACrefauthors}%
Pavlov, A\BPBI A.%
, Hurtgen, M\BPBI T.%
, Kasting, J\BPBI F.%
\BCBL {}\ \BBA {} Arthur, M\BPBI A.%
\end{APACrefauthors}%
\unskip\
\newblock
\APACrefYearMonthDay{2003}{}{}.
\newblock
{\BBOQ}\APACrefatitle {Methane-rich {Proterozoic} atmosphere?} {Methane-rich
  {Proterozoic} atmosphere?}{\BBCQ}
\newblock
\APACjournalVolNumPages{Geology}{31}{1}{87--90}.
\newblock
\begin{APACrefDOI} \doi{10.1130/0091-7613(2003)031<0087:MRPA>2.0.CO;2}
  \end{APACrefDOI}
\PrintBackRefs{\CurrentBib}

\bibitem [\protect \citeauthoryear {%
Pierce%
\ \BBA {} Adams%
}{%
Pierce%
\ \BBA {} Adams%
}{%
{\protect \APACyear {2006}}%
}]{%
pierce_global_2006}
\APACinsertmetastar {%
pierce_global_2006}%
\begin{APACrefauthors}%
Pierce, J\BPBI R.%
\BCBT {}\ \BBA {} Adams, P\BPBI J.%
\end{APACrefauthors}%
\unskip\
\newblock
\APACrefYearMonthDay{2006}{}{}.
\newblock
{\BBOQ}\APACrefatitle {Global evaluation of {CCN} formation by direct emission
  of sea salt and growth of ultrafine sea salt} {Global evaluation of {CCN}
  formation by direct emission of sea salt and growth of ultrafine sea
  salt}.{\BBCQ}
\newblock
\APACjournalVolNumPages{Journal of Geophysical Research}{111}{D6}{D06203}.
\newblock
\begin{APACrefDOI} \doi{10.1029/2005JD006186} \end{APACrefDOI}
\PrintBackRefs{\CurrentBib}

\bibitem [\protect \citeauthoryear {%
Pope%
, Bird%
\BCBL {}\ \BBA {} Rosing%
}{%
Pope%
\ \protect \BOthers {.}}{%
{\protect \APACyear {2012}}%
}]{%
pope_isotope_2012}
\APACinsertmetastar {%
pope_isotope_2012}%
\begin{APACrefauthors}%
Pope, E\BPBI C.%
, Bird, D\BPBI K.%
\BCBL {}\ \BBA {} Rosing, M\BPBI T.%
\end{APACrefauthors}%
\unskip\
\newblock
\APACrefYearMonthDay{2012}{{\APACmonth{03}}}{}.
\newblock
{\BBOQ}\APACrefatitle {Isotope composition and volume of {Earth}'s early
  oceans} {Isotope composition and volume of {Earth}'s early oceans}.{\BBCQ}
\newblock
\APACjournalVolNumPages{Proceedings of the National Academy of
  Sciences}{109}{12}{4371--4376}.
\newblock
\begin{APACrefDOI} \doi{10.1073/pnas.1115705109} \end{APACrefDOI}
\PrintBackRefs{\CurrentBib}

\bibitem [\protect \citeauthoryear {%
Prijith%
, Aloysius%
\BCBL {}\ \BBA {} Mohan%
}{%
Prijith%
\ \protect \BOthers {.}}{%
{\protect \APACyear {2014}}%
}]{%
prijith_relationship_2014}
\APACinsertmetastar {%
prijith_relationship_2014}%
\begin{APACrefauthors}%
Prijith, S.%
, Aloysius, M.%
\BCBL {}\ \BBA {} Mohan, M.%
\end{APACrefauthors}%
\unskip\
\newblock
\APACrefYearMonthDay{2014}{{\APACmonth{02}}}{}.
\newblock
{\BBOQ}\APACrefatitle {Relationship between wind speed and sea salt aerosol
  production: {A} new approach} {Relationship between wind speed and sea salt
  aerosol production: {A} new approach}.{\BBCQ}
\newblock
\APACjournalVolNumPages{Journal of Atmospheric and Solar-Terrestrial
  Physics}{108}{}{34--40}.
\newblock
\begin{APACrefDOI} \doi{10.1016/j.jastp.2013.12.009} \end{APACrefDOI}
\PrintBackRefs{\CurrentBib}

\bibitem [\protect \citeauthoryear {%
Reinhard%
\ \protect \BOthers {.}}{%
Reinhard%
\ \protect \BOthers {.}}{%
{\protect \APACyear {2020}}%
}]{%
reinhard_oceanic_2020}
\APACinsertmetastar {%
reinhard_oceanic_2020}%
\begin{APACrefauthors}%
Reinhard, C\BPBI T.%
, Olson, S\BPBI L.%
, Kirtland~Turner, S.%
, Pälike, C.%
, Kanzaki, Y.%
\BCBL {}\ \BBA {} Ridgwell, A.%
\end{APACrefauthors}%
\unskip\
\newblock
\APACrefYearMonthDay{2020}{{\APACmonth{11}}}{}.
\newblock
{\BBOQ}\APACrefatitle {Oceanic and atmospheric methane cycling in the {cGENIE}
  {Earth} system model – release v0.9.14} {Oceanic and atmospheric methane
  cycling in the {cGENIE} {Earth} system model – release v0.9.14}.{\BBCQ}
\newblock
\APACjournalVolNumPages{Geoscientific Model Development}{13}{11}{5687--5706}.
\newblock
\begin{APACrefDOI} \doi{10.5194/gmd-13-5687-2020} \end{APACrefDOI}
\PrintBackRefs{\CurrentBib}

\bibitem [\protect \citeauthoryear {%
Rose%
}{%
Rose%
}{%
{\protect \APACyear {2015}}%
}]{%
rose_stable_2015}
\APACinsertmetastar {%
rose_stable_2015}%
\begin{APACrefauthors}%
Rose, B\BPBI E\BPBI J.%
\end{APACrefauthors}%
\unskip\
\newblock
\APACrefYearMonthDay{2015}{{\APACmonth{02}}}{}.
\newblock
{\BBOQ}\APACrefatitle {Stable “{Waterbelt}” climates controlled by tropical
  ocean heat transport: {A} nonlinear coupled climate mechanism of relevance to
  {Snowball} {Earth}} {Stable “{Waterbelt}” climates controlled by tropical
  ocean heat transport: {A} nonlinear coupled climate mechanism of relevance to
  {Snowball} {Earth}}.{\BBCQ}
\newblock
\APACjournalVolNumPages{Journal of Geophysical Research:
  Atmospheres}{120}{4}{1404--1423}.
\newblock
\begin{APACrefDOI} \doi{10.1002/2014JD022659} \end{APACrefDOI}
\PrintBackRefs{\CurrentBib}

\bibitem [\protect \citeauthoryear {%
Som%
\ \protect \BOthers {.}}{%
Som%
\ \protect \BOthers {.}}{%
{\protect \APACyear {2016}}%
}]{%
som_earths_2016}
\APACinsertmetastar {%
som_earths_2016}%
\begin{APACrefauthors}%
Som, S\BPBI M.%
, Buick, R.%
, Hagadorn, J\BPBI W.%
, Blake, T\BPBI S.%
, Perreault, J\BPBI M.%
, Harnmeijer, J\BPBI P.%
\BCBL {}\ \BBA {} Catling, D\BPBI C.%
\end{APACrefauthors}%
\unskip\
\newblock
\APACrefYearMonthDay{2016}{}{}.
\newblock
{\BBOQ}\APACrefatitle {Earth's air pressure 2.7 billion years ago constrained
  to less than half of modern levels} {Earth's air pressure 2.7 billion years
  ago constrained to less than half of modern levels}.{\BBCQ}
\newblock
\APACjournalVolNumPages{Nature Geoscience}{9}{6}{448--+}.
\newblock
\begin{APACrefDOI} \doi{10.1038/NGEO2713} \end{APACrefDOI}
\PrintBackRefs{\CurrentBib}

\bibitem [\protect \citeauthoryear {%
Spalding%
\ \BBA {} Fischer%
}{%
Spalding%
\ \BBA {} Fischer%
}{%
{\protect \APACyear {2019}}%
}]{%
spalding_shorter_2019}
\APACinsertmetastar {%
spalding_shorter_2019}%
\begin{APACrefauthors}%
Spalding, C.%
\BCBT {}\ \BBA {} Fischer, W\BPBI W.%
\end{APACrefauthors}%
\unskip\
\newblock
\APACrefYearMonthDay{2019}{{\APACmonth{05}}}{}.
\newblock
{\BBOQ}\APACrefatitle {A shorter {Archean} day-length biases interpretations of
  the early {Earth}'s climate} {A shorter {Archean} day-length biases
  interpretations of the early {Earth}'s climate}.{\BBCQ}
\newblock
\APACjournalVolNumPages{Earth and Planetary Science Letters}{514}{}{28--36}.
\newblock
\begin{APACrefDOI} \doi{10.1016/j.epsl.2019.02.032} \end{APACrefDOI}
\PrintBackRefs{\CurrentBib}

\bibitem [\protect \citeauthoryear {%
Stanhill%
}{%
Stanhill%
}{%
{\protect \APACyear {1994}}%
}]{%
stanhill_changes_1994}
\APACinsertmetastar {%
stanhill_changes_1994}%
\begin{APACrefauthors}%
Stanhill, G.%
\end{APACrefauthors}%
\unskip\
\newblock
\APACrefYearMonthDay{1994}{{\APACmonth{05}}}{}.
\newblock
{\BBOQ}\APACrefatitle {Changes in the rate of evaporation from the dead sea}
  {Changes in the rate of evaporation from the dead sea}.{\BBCQ}
\newblock
\APACjournalVolNumPages{International Journal of Climatology}{14}{4}{465--471}.
\newblock
\begin{APACrefDOI} \doi{10.1002/joc.3370140409} \end{APACrefDOI}
\PrintBackRefs{\CurrentBib}

\bibitem [\protect \citeauthoryear {%
Twomey%
}{%
Twomey%
}{%
{\protect \APACyear {1974}}%
}]{%
twomey_pollution_1974}
\APACinsertmetastar {%
twomey_pollution_1974}%
\begin{APACrefauthors}%
Twomey, S.%
\end{APACrefauthors}%
\unskip\
\newblock
\APACrefYearMonthDay{1974}{{\APACmonth{12}}}{}.
\newblock
{\BBOQ}\APACrefatitle {Pollution and the planetary albedo} {Pollution and the
  planetary albedo}.{\BBCQ}
\newblock
\APACjournalVolNumPages{Atmospheric Environment (1967)}{8}{12}{1251--1256}.
\newblock
\begin{APACrefDOI} \doi{10.1016/0004-6981(74)90004-3} \end{APACrefDOI}
\PrintBackRefs{\CurrentBib}

\bibitem [\protect \citeauthoryear {%
Way%
\ \protect \BOthers {.}}{%
Way%
\ \protect \BOthers {.}}{%
{\protect \APACyear {2017}}%
}]{%
way_resolving_2017}
\APACinsertmetastar {%
way_resolving_2017}%
\begin{APACrefauthors}%
Way, M\BPBI J.%
, Aleinov, I.%
, Amundsen, D\BPBI S.%
, Chandler, M\BPBI A.%
, Clune, T\BPBI L.%
, Genio, A\BPBI D\BPBI D.%
\BDBL {}Tsigaridis, K.%
\end{APACrefauthors}%
\unskip\
\newblock
\APACrefYearMonthDay{2017}{{\APACmonth{07}}}{}.
\newblock
{\BBOQ}\APACrefatitle {Resolving {Orbital} and {Climate} {Keys} of {Earth} and
  {Extraterrestrial} {Environments} with {Dynamics} ({ROCKE}-{3D}) 1.0: {A}
  {General} {Circulation} {Model} for {Simulating} the {Climates} of {Rocky}
  {Planets}} {Resolving {Orbital} and {Climate} {Keys} of {Earth} and
  {Extraterrestrial} {Environments} with {Dynamics} ({ROCKE}-{3D}) 1.0: {A}
  {General} {Circulation} {Model} for {Simulating} the {Climates} of {Rocky}
  {Planets}}.{\BBCQ}
\newblock
\APACjournalVolNumPages{The Astrophysical Journal Supplement
  Series}{231}{1}{12}.
\newblock
\begin{APACrefDOI} \doi{10.3847/1538-4365/aa7a06} \end{APACrefDOI}
\PrintBackRefs{\CurrentBib}

\bibitem [\protect \citeauthoryear {%
Williams%
}{%
Williams%
}{%
{\protect \APACyear {2000}}%
}]{%
williams_geological_2000}
\APACinsertmetastar {%
williams_geological_2000}%
\begin{APACrefauthors}%
Williams, G.%
\end{APACrefauthors}%
\unskip\
\newblock
\APACrefYearMonthDay{2000}{}{}.
\newblock
{\BBOQ}\APACrefatitle {Geological constraints on the {Precambrian} history of
  {Earth}'s rotation and the {Moon}'s orbit} {Geological constraints on the
  {Precambrian} history of {Earth}'s rotation and the {Moon}'s orbit}.{\BBCQ}
\newblock
\APACjournalVolNumPages{Reviews of Geophysics}{38}{}{37--59}.
\PrintBackRefs{\CurrentBib}

\end{thebibliography}

%%%%%%%%%%%%%%%%%%%%%%%%%%%%%%%%
%% Optional Appendix goes here
%

\appendix
\section{Supporting Information}
This supporting information includes three figures that highlight key aspects of the ocean circulation in our present-day Earth configuration. These figures serve to (1) illustrate that our model reproduces key aspects of the Atlantic Meridional Overturning Circulation (AMOC) despite our assumption of a relatively shallow ocean to save computational expense, (2) demonstrate that AMOC and associated ocean heat transport increase with increasing salinity, and (3) support our conclusion that dynamical effects are the primary contributor to the climate sensitivity to salinity in Figures 1 and 2 in the main text. 
\clearpage

 \begin{figure}
\noindent\includegraphics[width=\textwidth]{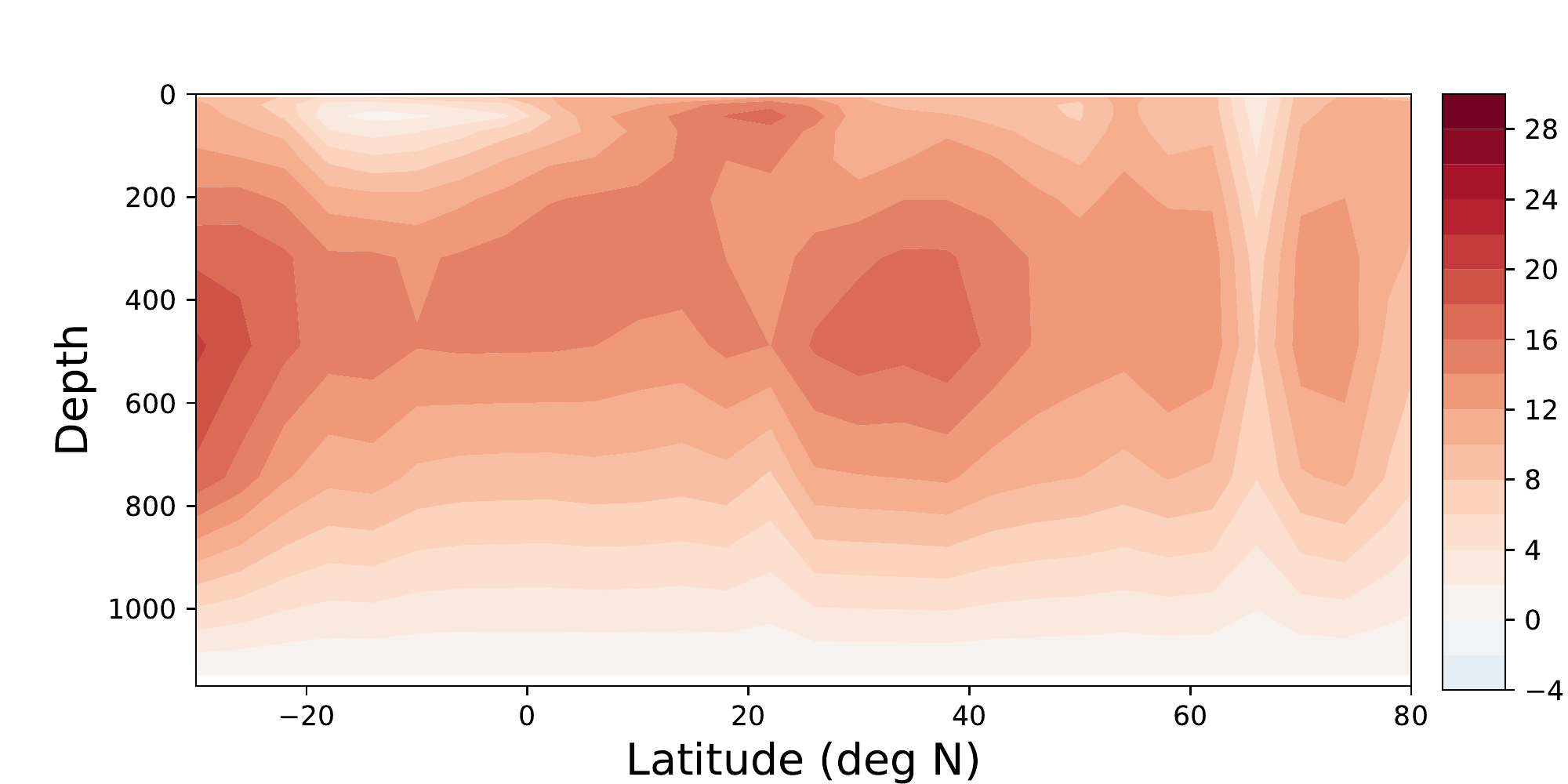}
 \caption{Atlantic streamfunction for our present-day Earth scenario. Our model qualitatively reproduces the AMOC, despite our assumption of a shallow, flat-bottom ocean.}
\end{figure}

\clearpage

 \begin{figure}
\noindent\includegraphics[width=\textwidth]{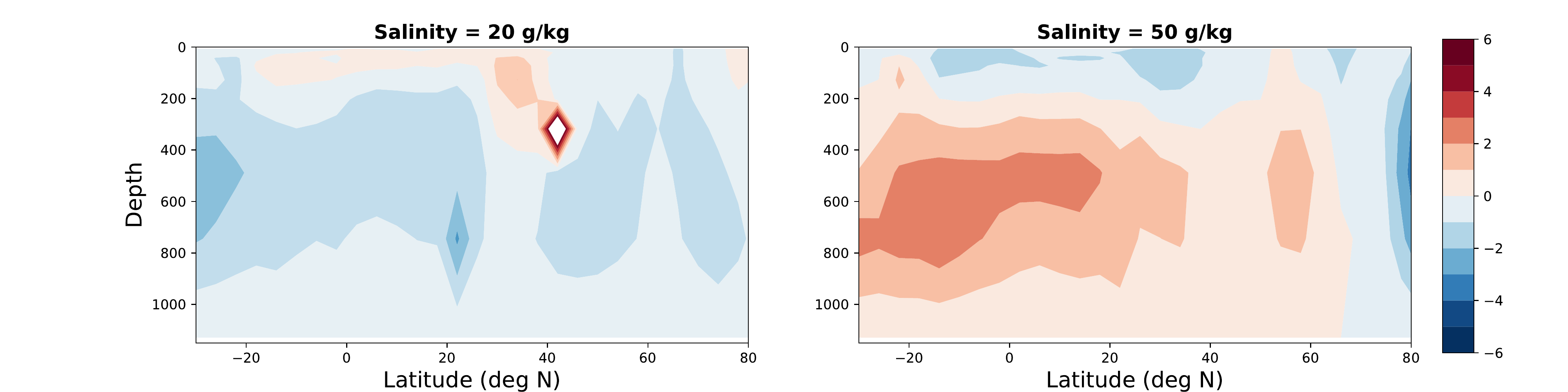}
 \caption{Atlantic streamfunction sensitivity to salinity in our present-day simulations. Shown are the differences between the streamfunctions for our 20 g/kg (left) and 50 g/kg (right) salinity scenarios and the streamfunction for our present-day salinity scenario (Figure A1) in units of Sv.}
\end{figure}

\clearpage

 \begin{figure}
\noindent\includegraphics[width=\textwidth]{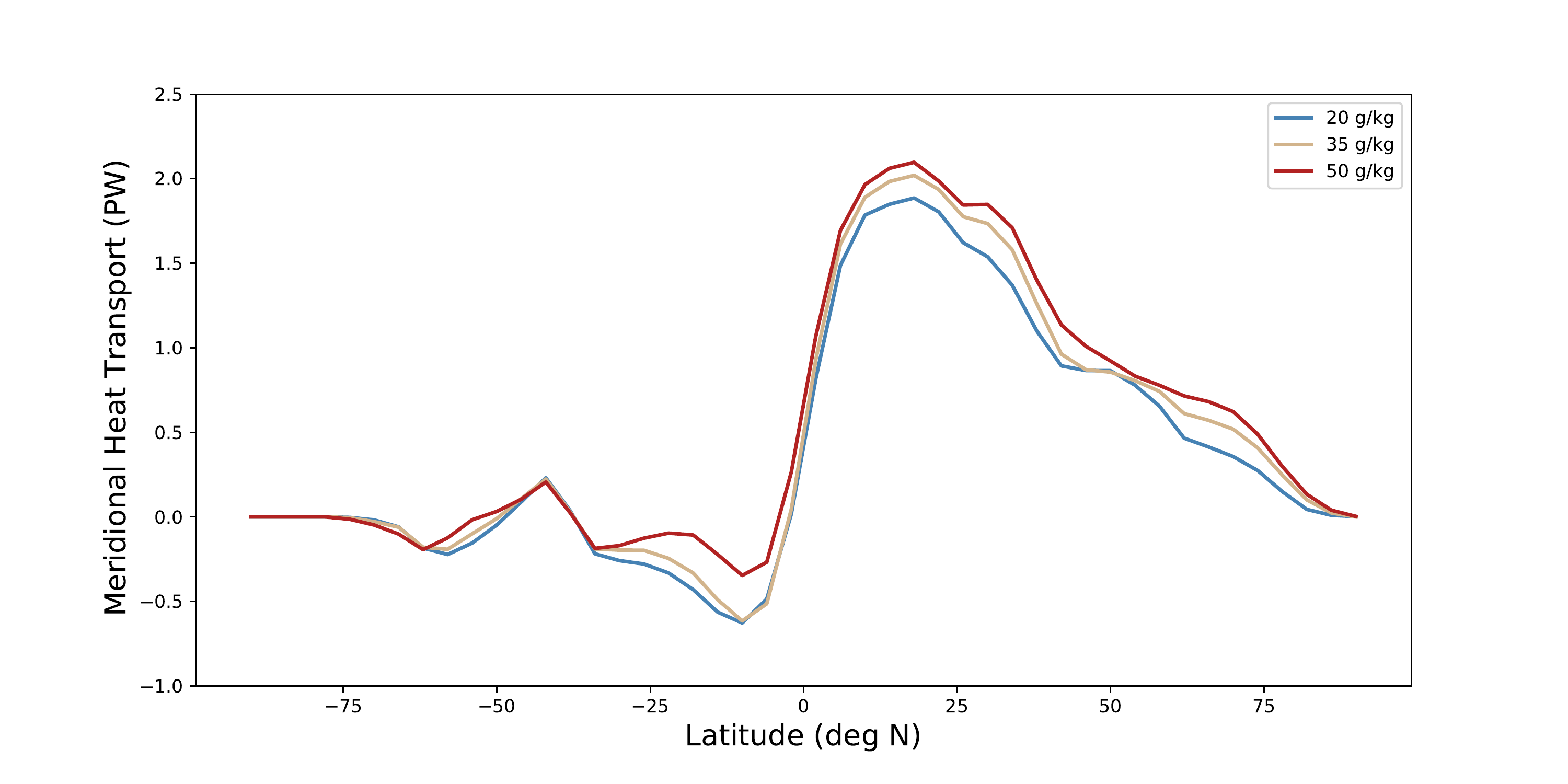}
 \caption{Meridional ocean heat transport in our present-day simulations. The colored lines correspond to salinities of 20 g/kg (blue), 35 g/kg (tan), and 50 g/kg (red). Heat transport to the northern high latitudes increase with ocean salinity, but this effect is muted in the southern hemisphere. This asymmetry is also reflected in the sensitivity of surface temperature and sea ice cover to ocean salinity (Figures 1, 2).}
\end{figure}

\clearpage

 \begin{figure}
\noindent\includegraphics[width=\textwidth]{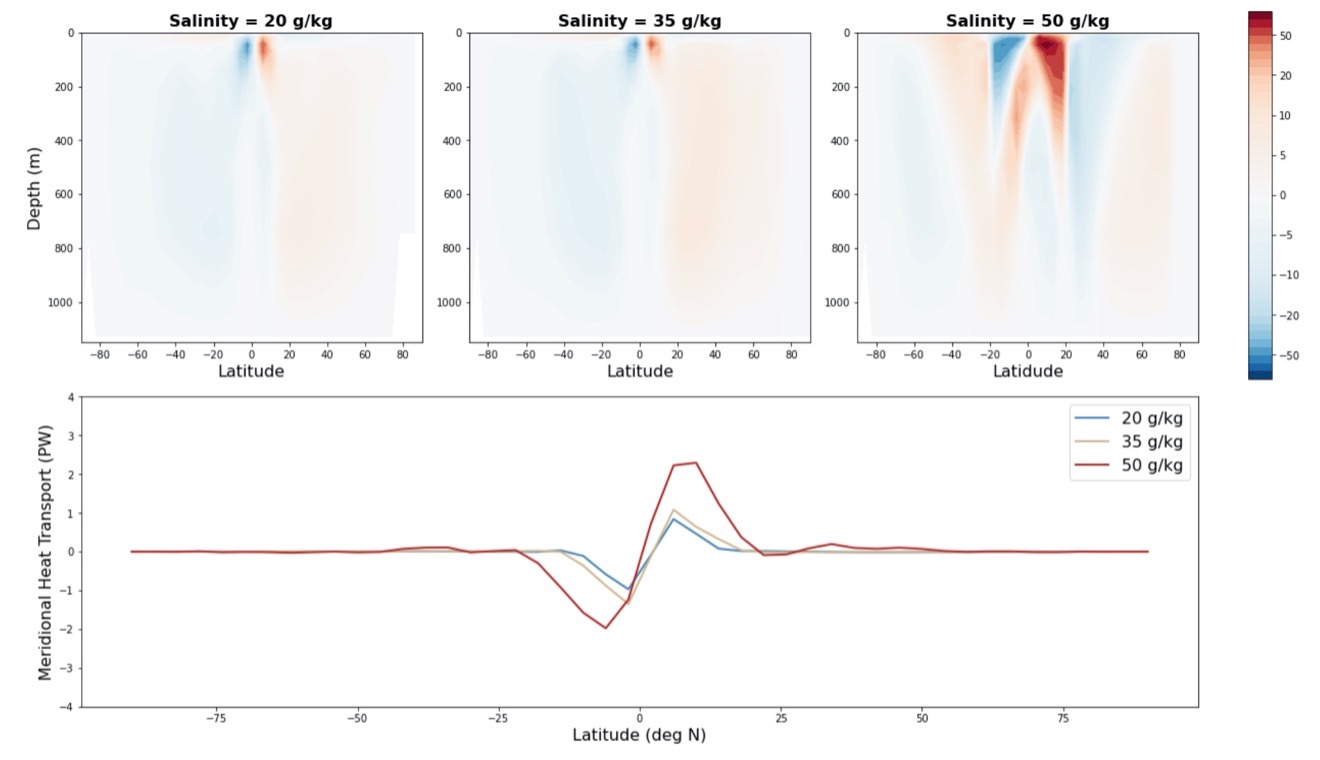}
 \caption{Global streamfunction (top) and meridional ocean heat transport (bottom) in select Archean scenarios. Shown here are Archean simulations with differing salinities and 60x PIL pCO$_2$, corresponding to the middle row of Figure 4 in the main text. Note that the 20 g/kg and 35 g/kg salinity scenarios (top left and middle panels; tan and blue lines in bottom panel) are water belt states whereas the higher salinity scenario has considerably more open water (see Figure 4), complicating direct comparison and interpretation. Ocean circulation and associated heat transport are influenced by both salinity and the position of the ice line, introducing the possibility of feedbacks between ocean circulation, ice extent, and surface temperature that may amplify the direct effects of salinity. }
\end{figure}

\end{document}